\documentclass[preprint,12pt]{elsarticle}

\usepackage{amssymb}

\journal{Nuclear Physics A}

\begin{document}

\begin{frontmatter}



\title{Photon production in Pb+Pb collisions at $\sqrt{s_{NN}}$=2.76 TeV}


\author{Yong-Ping Fu, Qin Xi}

\address{Department of Physics and mathematics, Dianxi Science and Technology Normal University, Lincang 677000, China}

\begin{abstract}
We calculate the high energy photon production from the Pb+Pb collisions for different centrality classes at $\sqrt{s_{NN}}$=2.76 TeV Large Hadron Collider (LHC) energy. The jet energy loss in the jet fragmentation, jet-photon conversion and jet bremsstrahlung is considered by using the Wang-Huang-Sarcevic (WHS) and Baier-Dokshitzer-Mueller-Peigne-Schiff (BDMPS) models. We use the (1+1)-dimensional ideal relativistic hydrodynamics to study the collective transverse flow and space-time evolution of the quark gluon plasma (QGP). The numerical results agree well with the ALICE data of the direct photons from the Pb+Pb collisions ($\sqrt{s_{NN}}$=2.76 TeV) for 0-20\%, 20-40\% and 40-80\% centrality classes.

\end{abstract}


\end{frontmatter}



\section{Introduction}

The electromagnetic radiation produced from the relativistic heavy-ion collisions is considered to be a useful probe for the investigation of the perturbative QCD
(pQCD) and QGP. The photons do not participate in the strong interaction directly, and the mean-free path of the photons is larger than the collision system. Thus the photons can escape to the detector almost undistorted through the strongly interacting system.
The photon production can test the predictions of pQCD
calculations, and probe the properties of the QGP.

The data of the direct photon transverse momentum spectrum in Pb+Pb collisions at $\sqrt{s_{NN}}$=2.76 TeV for the 0-20\%, 20-40\% and 40-80\% centrality classes is presented by ALICE Collaboration \cite{ALICE 1}. The photons are produced from various
processes in relativistic heavy-ion
collisions: the primary hard photons from the initial parton collisions and the jet fragmentation
\cite{hard photon 1,hard photon 2,hard photon 3}, the thermal photons from the pre-equilibrium hot matter, equilibrium QGP
\cite{th photon 1,th photon 2,th photon 3,new photon,photon 1} and hadronic gas (HG)
\cite{th photon 2,HG photon 1,HG photon 2}, the photons from the jet-photon conversion and jet bremsstrahlung in
the thermal medium \cite{jet-photon 1,jet-photon 2,jet-photon 3,jet-photon 4,jet-photon 4.1,jet-brem 1,jet-brem 2}, and the photons from the hadronic decays after the freeze-out \cite{decay 1,decay 2,decay 3}. The local thermalization is not reached immediately in relativistic heavy-ion collisions. The pre-equilibrium models in the early stages of the collisions suggest that local thermalization is reached at the time of 0.1-2 fm/c \cite{pre-equi}. The pre-equilibrium photons may not be entirely negligible at LHC. The largest contributions of hadron decays come from $\pi^{0}$(80-85\%), $\eta$, and $\omega$ decays. The photons from the hadron decays after the freeze-out have already been excluded by ALICE experiments.

In the present work we investigate the photons production from Pb+Pb collisions at $\sqrt{s_{NN}}$=2.76 TeV, and compare with the ALICE experimental data. The annihilation and Compton scattering of initial partons, and jet fragmentation are well-known dominant sources of large transverse momentum ($p_{T}$) photons in relativistic heavy-ion collisions. The thermal photons from the QGP and HG are dominant in the relatively small $p_{T}$ region. Furthermore, the electromagnetic radiation from jet-plasma interaction is also an important source of large $p_{T}$ photons. Fries \emph{et al}. \cite{jet-photon 1,jet-photon 2} have only studied the quark jet-photon conversion in the plasma, we evaluate the photon production from the gluon jet-photon conversion in this paper. The centrality dependence of the photon spectra is scaled by the number of binary collisions ($N_{\mathrm{coll}}$), the number of participants ($N_{\mathrm{par}}$), and the hadron multiplicity $dN/dy$.

The relativistic hydrodynamics can describe the collective flow of the strongly interacting matter produced from the relativistic heavy-ion collisions. The (0+1)-dimensional (1-D) Bjorken hydrodynamical equations provide a good estimate for the longitudinal expansion of the QGP \cite{Bjorken}. However the transverse expansion becomes important at LHC energies due to the large life time of the QGP. The transverse flow effects have been studied by the (1+1)-D relativistic hydrodynamics which assumes cylindrical symmetry along the transverse direction and boost invariant along the longitudinal direction \cite{1+3D1,1+3D2,1+3D3,1+3D4}. In this paper, we study the effect of the collective transverse flow in photon production from the QGP, HG and jet-medium interactions. We find that the transverse expansion leads to a rapid cooling of the fire ball. The thermal photon spectrum will be suppressed by the transverse flow effect at small transverse momentum region due to the decreasing of the medium life time.

In the jet-medium interactions the energetic partons will lose their energies. For high energy partons, the medium induced energy loss is dominated by the gluon bremsstrahlung \cite{Eloss1}. In Ref. \cite{jet-photon 3,jet-photon 4,jet-photon 4.1} the authors use the Arnold-Moore-Yaffe (AMY) formalism to investigate the energy loss of jet-plasma interactions. The AMY model assumes that high energy partons evolve in the QGP according to the Fokker-Planck rate equations for their momentum distributions. The energy loss is described as a dependence of the parton momentum distribution $dN^{jet}/(dp^{jet}_{T}dy_{jet})$ on time. In this paper we use the WHS model \cite{WHS1,WHS2,WHS3,WHS4} to calculate the energy loss on jet fragmentation, jet-photon conversion, and the jet bremsstrahlung. The WHS model assumes that the probability for a parton to scatter $n$ times within distance $L$ is given by the Poisson distribution. The energy loss due to the inelastic scatterings is determined by the energy loss per unit distance $dE_{a}/dx$. The parton momentum distribution $dN^{jet}/(dp^{jet}_{T}dy_{jet})$ with energy loss depends on the distance of jet passing through the hot medium in the WHS approach. However an average energy loss per unit distance was used in Refs. \cite{WHS1,WHS2,WHS3,WHS4}. In the present work we use the rigorous formalism of energy loss $dE_{a}/dx$ developed by the BDMPS model \cite{BDMPS1,BDMPS2} instead of the average energy loss. We have studied the effect of the jet energy loss in the jet-dilepton conversion by using the WHS and BDMPS frameworks. The numerical results of the jet energy loss agree with the results from the AMY approach \cite{fu}.

The paper is organized as follows. In Sec.II we present the
production of prompt photons in relativistic heavy-ion
collisions. The annihilation, Compton and fragmentation processes are
presented. In Sec.III we discuss the initial conditions of the QGP and the production of thermal
photons. In Sec.IV the jet-medium interaction of the jet-photon conversion and the jet bremsstrahlung is discussed. The numerical results
of the photon production are plotted in Sec.V. Finally, the conclusion
is given in Sec.VI.

\section{Prompt photon production}

\subsection{Direct production}

In relativistic heavy-ion collisions, the initial parton collisions will produce large transverse momentum ($p_{T}$) photons through quark-antiquark
annihilation, Compton scattering and gluon-photon coupling. In addition, the final state jets of parton collisions can fragment into photons.
The high energy jets will loss their energies before they fragment into photons. Prompt photons are defined as the photons produced from the direct and fragmentation processes.

The large $p_{T}$ photons can be directly produced by the quark-antiquark annihilation ($q\bar{q}\rightarrow
g\gamma$, $q\bar{q}\rightarrow
\gamma\gamma$), Compton collision ($qg\rightarrow q
\gamma$) and gluon-photon coupling ($gg\rightarrow g\gamma$, $gg\rightarrow \gamma\gamma$) in the hadronic
collisions($AB\rightarrow \gamma(\gamma\gamma) X$). The invariant cross section from above processes is
\begin{eqnarray}
\frac{d\sigma_{\mathrm{dir}}}{d^{2}p_{T}dy}\label{dir}&=&\sum_{a,b}\int^{1}_{x^{min}_{a}}
dx_{a}f_{a}^{A}(x_{a},Q^{2})f_{b}^{B}(x_{b},Q^{2})
\nonumber\\[1mm]
&&\times
\frac{x_{a}x_{b}}{\pi(x_{a}-x_{1})}K_{\mathrm{dir}}\frac{d\hat{\sigma}_{ab\rightarrow \gamma d}}{d\hat{t}}(\hat{s},\hat{u},\hat{t}),
\end{eqnarray}
here $f^{A}_{a}(x_{a},Q^{2})$ and
$f^{B}_{b}(x_{b},Q^{2})$ are parton distributions of nucleus, $x_{a}$
and $x_{b}$ are the parton's momentum fraction, $x^{min}_{a}=x_{1}/(1-x_{2})$ and
$x_{b}=x_{a}x_{2}/(x_{a}-x_{1})$. The variables are
$x_{1}=x_{T}e^{y}/2$, $x_{2}=x_{T}e^{-y}/2$,
$x_{T}=2p_{T}/\sqrt{s_{NN}}$. $y$ is the
rapidity and $\sqrt{s_{NN}}$ is the center-of-mass energy of the colliding nucleus per nucleon pair.
The sum runs over the $u$, $d$, $s$ (anti)quarks and gluon. The Mandelstam variables $\hat{s}$, $\hat{u}$, $\hat{t}$ of the the subprocesses $d\hat{\sigma}/d\hat{t}[ab\rightarrow \gamma(\gamma\gamma) d]$ \cite{hard photon 2,hard photon 3} are $\hat{s}=x_{a}x_{b}s_{NN}$, $\hat{u}=-s_{NN}x_{b}x_{1}$ and $\hat{t}=-s_{NN}x_{a}x_{2}$. The factor $K_{\mathrm{dir}}$=1.5 is used to account for the next-to-leading order (NLO) corrections \cite{jet-photon 3}.

The parton distribution $f_{i}^{A}(x_{i},Q^{2})$ ($i=a,b$) of the nucleus is
given by
\begin{eqnarray}
f_{i}^{A}(x_{i},Q^{2})\!\!=\!\!R_{i}^{A}(x_{i},Q^{2})
\!\!\left[\!\frac{Z}{A}f^{p}_{i}(x_{i},Q^{2})\!\!+\!\!\frac{N}{A}f^{n}_{i}(x_{i},Q^{2})\right]\!\!,
\end{eqnarray}
where $R^{A}_{i}(x_{i},Q^{2})$ is the nuclear modification factor, $Z$ is the proton number, $N$ is the neutron number and
$A$ is the nucleon number. $f^{p}_{i}(x_{i},Q^{2})$ and
$f^{n}_{i}(x_{i},Q^{2})$ are the parton distributions of protons and
neutrons, respectively. We choose the momentum scale as
$Q^{2}=4p_{T}^{2}$. We use the CTEQ6L1 parton distributions \cite{CTEQ6M 1} and EPS09 nuclear modifications \cite{EPS09 1} in the Pb+Pb collisions.

The QCD annihilation and Compton processes [$O(\alpha\alpha_{s})$] is the dominant contribution of the direct photon production. The photon production from the QED annihilation [$O(\alpha^{2})$] and gluon-photon coupling [$O(\alpha\alpha_{s}^{3})$ and $O(\alpha^{2}\alpha_{s}^{2})$] is suppressed by the  higher order electromagnetic coupling parameter $\alpha$ and strong coupling parameter $\alpha_{s}$. However, because of the abundance of gluons at low $x_{i}$ the gluon-photon coupling contributions are not negligible at small $p_{T}$ region \cite{hard photon 2}.

\subsection{Jet fragmentation}

The invariant cross section for the $AB\rightarrow (c\rightarrow\gamma)X$ interaction arising from the single bremsstrahlung in the vacuum can be written as
\begin{eqnarray}\label{frag}
\frac{d\sigma_{\mathrm{fra}}}{d^{2}p_{T}dy}&\!\!=&\!\!\sum_{a,b}\int^{1}_{x_{a}^{min}}\!\!
dx_{a}\int^{1}_{x_{b}^{min}}\!\! dx_{b} f_{a}^{A}(x_{a},Q^{2}) f_{b}^{B}(x_{b},Q^{2})
\nonumber\\[1mm]
&&\times
D_{c}^{\gamma}(z_{c},Q^{2})\frac{1}{\pi z_{c}}K_{\mathrm{fra}}
\frac{d\hat{\sigma}_{ab\rightarrow c d}}{d\hat{t}}(\hat{s},\hat{u},\hat{t}),
\end{eqnarray} 
where $x_{a}^{min}=x_{1}/(1-x_{2})$, $x_{b}^{min}=x_{a}x_{2}/(x_{a}-x_{1})$. $z_{c}=(x_{a}x_{2}+x_{b}x_{1})/x_{a}x_{b}$ is the
momentum fraction of the final states. The subprocesses of the photon fragmentation are $qq'\rightarrow qq'$, $q\bar{q}'\rightarrow q\bar{q}'$,
$qq\rightarrow qq$, $q\bar{q}\rightarrow q'\bar{q}'$,
$q\bar{q}\rightarrow q\bar{q}$, $gg\rightarrow q\bar{q}$,
$qg\rightarrow qg$, $q\bar{q}\rightarrow gg$ and $gg\rightarrow gg$ \cite{hard photon 3}. The Mandelstam variables $\hat{s}$, $\hat{u}$, $\hat{t}$ of the the differential cross sections $d\hat{\sigma}/d\hat{t}(ab\rightarrow c d)$ are $\hat{s}=x_{a}x_{b}s_{NN}$, $\hat{u}=-s_{NN}x_{b}x_{1}/z_{c}$ and $\hat{t}=-s_{NN}x_{a}x_{2}/z_{c}$. The NLO correction factor of the fragmentation process is $K_{\mathrm{fra}}$=1.4 for the LHC energy \cite{jet-photon 3}.

The photon fragmentation function $D_{q}^{\gamma}$ and $D_{g}^{\gamma}$ contains an electromagnetic coupling parameter, so the bremsstrahlung  processes $ab\rightarrow (c\rightarrow \gamma)(d\rightarrow \gamma)$ and $ab\rightarrow (c\rightarrow \gamma)\gamma$ are negligible. If an energetic jet paces through a distance $L$ in the QGP, and fragments outside the system, the jet will loss its energy. Induced gluon bremsstrahlung, rather than elastic scattering of partons, is the dominant contribution of the jet energy loss

The prompt photon yield for different centrality classes is obtained by the following \cite{TAA 1}
\begin{eqnarray}
\frac{dN_{\mathrm{prom}}}{d^{2}p_{T}dy}(b)=\frac{\langle N_{\mathrm{coll}}\rangle_{b}}{\sigma_{\mathrm{inel}}^{NN}}\left(\frac{d\sigma_{\mathrm{dir}}}{d^{2}p_{T}dy}+\frac{d\sigma_{\mathrm{fra}}}{d^{2}p_{T}dy}\right),
\end{eqnarray}
where $\langle N_{\mathrm{coll}}\rangle_{b}$ is the average number of binary nucleon-nucleon collisions ($\langle N_{\mathrm{coll}}\rangle_{b}$=1210, 438, and 77 for the 0-20\%, 20-40\%, and 40-80\% class, respectively), and $\sigma_{\mathrm{inel}}^{NN}$=64 mb is the inelastic nucleon-nucleon cross section in Pb+Pb collisions at $\sqrt{s_{NN}}$=2.76 TeV \cite{ALICE 1,TAA 2}.

\subsection{Jet energy loss}
The energy loss of jets crossing the hot and dense plasma by means of the spectrum of energy loss per unit distance $dE_{a}/dx$ is determined by the BDMPS model \cite{BDMPS1,BDMPS2}
\begin{eqnarray}
\frac{dE_{a}}{dx}=\frac{\alpha_{s}c_{a}\mu_{\mathrm{D}}^{2}}{8\lambda_{g}}L\ln\frac{L}{\lambda_{g}},
\end{eqnarray}
where $c_{a}$=4/3 for quarks and 3 for gluon, $\mu^{2}_{\mathrm{D}}=4\pi\alpha_{s}T^{2}$, $\mu_{\mathrm{D}}$ is the Debye mass of the medium, $T$ is the temperature of the QGP. $\lambda_{g}=\pi\mu^{2}/\left[126\alpha_{s}^{2}\zeta(3)T^{3}\right]$ and $\lambda_{q}=9\lambda_{g}/4$ is the gluon and quark mean-free path, respectively \cite{Eloss1}. When an energetic parton is propagating through a QGP, the total energy loss is
\begin{eqnarray}
\triangle E_{a}
=\int_{0}^{L}\frac{dE_{a}}{dx} dx.
\end{eqnarray}

The effect of medium induced parton energy loss on jet fragmentation is presented by the WHS phenomenological model. This approach is useful for studies of the parton energy loss of fragmentation function and multiple final-state scatterings. The probability for a jet to catter $n$ times within a distance $L=n\lambda_{a}$ in the QGP is
\begin{eqnarray}
P_{a}(n)=\frac{( L/\lambda_{a})^{n}}{n!}e^{- L/\lambda_{a}},
\end{eqnarray}
here $\lambda_{a}$ is the mean-free path of the parton. If the average energy loss per scattering is $\varepsilon_{a}=\lambda_{a}\left(dE_{a}/dx\right)$, the modified photon fragmentation function can be written as
\begin{eqnarray}
D_{a}^{\gamma}(z_{a},Q^{2})=C_{n}\sum^{N}_{n=0}P_{a}(n)\frac{z_{a}^{n}}{z_{a}}D_{a}^{0\gamma}(z^{n}_{a},Q^{2}) ,
\end{eqnarray}
where $C_{n}=1/\left(\sum^{N}_{n=0}P_{a}(n)\right)$, $N=E^{\mathrm{jet}}_{T}/\varepsilon_{a}$ is the scattering number, $E^{\mathrm{jet}}_{T}=p^{\gamma}_{T}/z_{a}$ is the transverse energy of the jet, and $z^{n}_{a}=z_{a}/\left(1-\Delta E_{a}/E^{\mathrm{jet}}_{T} \right)$. We use the NLO parametrization for photon fragmentation functions $D_{q}^{0\gamma}$ and $D_{g}^{0\gamma}$ without the energy loss of the medium from Refs. \cite{photon frag 1} and \cite{photon frag 2}, respectively.

\section{Thermal photon production }

\subsection{Initial conditions and hydrodynamic equations }
We assume the hydrodynamic flow does not have transverse flow at the initial temperature $T_{0}$ and initial time $\tau_{0}$. $T_{0}$ and $\tau_{0}$ of the QGP can be related to the hadron multiplicity distribution in the final phase by \cite{1+3D2,condition 1}
\begin{eqnarray}
T^{3}_{0}\tau_{0}=\frac{2\pi^{4}}{45\zeta(3)}\frac{1}{4a_{Q}\pi R^{2}_{\perp 0}(b)}\frac{dN}{dy}(b),
\end{eqnarray}
where $a_{Q}=g_{Q}\pi^{2}/90$, and $g_{Q}=42.5$ for a system consisting of $u$, $d$, $s$ quarks and gluons. $b$ is the impact parameter. $R_{\perp 0}\approx1.2(N_{\mathrm{part}}/2)^{1/3}$ fm is the initial transverse radius of the system \cite{jet-photon 2}. $N_{\mathrm{part}}$ is the number of participant nucleons. $dN/dy$ is the hadron multiplicity. By using the Glauber simulation and the CMS experimental results \cite{condition 1,condition 2,condition 3,condition 4,condition 5}, the initial conditions are presented in the Table I. Since the emission of thermal photons is sensitive to the initial conditions, we fix the value of the initial time \cite{ALICE 1,condition 4} as $\tau_{0}$=0.15 fm/$c$.

\begin{table}
\centering\caption{Initial conditions of the hydrodynamical
expansion.}
\label{temperature}
\begin{tabular}{rcccccccc}
\hline
\hline & Centrality &$dN/dy$ &$N_{\mathrm{part}}$ &$T_{0}$(MeV) &$\tau_{0}$(fm/\emph{c})  & \\
\hline
&0-20\%  & 2117&$ 306 $&$ 762$&$ 0.15$&\\
&20-40\% & 958 &$ 156 $&$ 680$&$ 0.15$&\\
&40-80\% & 195 &$ 41  $&$ 538$&$ 0.15$&\\
\hline
\end{tabular}
\end{table}

The equation for conservation of energy-momentum of an ideal fluid produced in relativistic heavy-ion collisions is given by
\begin{eqnarray}
\partial_{\mu}\left[(\varepsilon+P)u^{\mu}u^{\nu}-Pg^{\mu\nu}\right]=0,
\label{conservation}
\end{eqnarray}
where $\varepsilon$ is the energy density, $P$ is the pressure. $u^{\mu}=\gamma_{r}(\tau,r)(t/\tau,v_{r}(\tau,r),z/\tau)$ is the velocity of the (1+1)-D fluid with cylindrical symmetry and boost invariant along the longitudinal direction \cite{1+3D1,1+3D3}, here $\gamma_{r}=\left[1-v_{r}^{2}(\tau,r)\right]^{-1/2}$, and $\tau=(t^{2}-z^{2})^{1/2}$.

We take the equations of state for the QGP phase and hadronic phase as \cite{th photon 7}
\begin{eqnarray}
\varepsilon_{Q}=g_{Q}\frac{\pi^{2}}{30}T^{4}+B,
\end{eqnarray}
\begin{eqnarray}
P_{Q}=g_{Q}\frac{\pi^{2}}{90}T^{4}-B,
\end{eqnarray}
\begin{eqnarray}
\varepsilon_{H}=g_{H}\frac{\pi^{2}}{30}T^{4},
\end{eqnarray}
\begin{eqnarray}
P_{H}=g_{H}\frac{\pi^{2}}{90}T^{4},
\end{eqnarray}
where $B$ is the bag constant. It is not proper to treat hadronic gas as a dilute gas consisting of only pions, so we use $g_{H}\approx$ 4.59 for the hadronic gas \cite{1+3D4}. The hydrodynamics equation Eq.(\ref{conservation}) for a transverse and longitudinal expansion can be written as \cite{fu}
\begin{eqnarray}
\frac{\partial \varepsilon}{\partial\tau}\!+\!\frac{\varepsilon\!\!+\!\!P}{\tau}\!+\!(\varepsilon\!\!+\!\!P)\!\!\left[\frac{\partial v_{r}(\tau,r)}{\partial r}+u^{\mu}\partial_{\mu}\ln \gamma_{r}(\tau,r)\right]\!=\!0.
\label{hydro}
\end{eqnarray}
If there is only a longitudinal expansion of the QGP ($v_{r}=0$), Eq.(\ref{hydro}) becomes the well-known Bjorken equation. The initial conditions of the transverse expansion are chosen such that $v_{r}(\tau_{0},r)$=0 along with a given initial temperature $T(\tau_{0},r)=T_{0}$ within the transverse radius \cite{1+3D4}.

\subsection{Thermal photons from QGP }
By using the hard thermal loop (HTL) approximation, the production rate of thermal photons produced from the annihilation and Compton interaction of thermal partons is given by the following
\cite{th photon 2,th photon 4}
\begin{eqnarray}
E_{\gamma}\frac{dR_{\mathrm{QGP}}}{d^{3}p}&=&\sum_{q}\left(\frac{e_{q}}{e}\right)^{2}\frac{\alpha\alpha_{s}T^{2}}{4\pi^{2}} f^{\mathrm{FD}}_{\mathrm{th}}(\emph{\textbf{p}}_{\gamma})
\nonumber\\[1mm]
&&\times
\left[2\mathrm{ln}\left(\frac{3E_{\gamma}}{\pi\alpha_{s}T}\right)\!+C_{\mathrm{Com}}\!+C_{\mathrm{ann}}\right],
\end{eqnarray}
where $q$=$u$, $d$, $s$ quarks, and $f^{\mathrm{FD}}_{\mathrm{th}}$ is the Fermi-Dirac distribution of thermal
partons. The parameters are $C_{\mathrm{Com}}$=-0.416 and $C_{\mathrm{ann}}$=-1.916. We use the temperature dependent coupling constant $\alpha_{s}(T)=6\pi/\left[(33-2N_{f})\ln(8T/T_{c})\right]$ from Ref.\cite{temp 1}, here the critical temperature $T_{c}$= 160 MeV, the flavor number of the quarks $N_{f}\approx$2.5 to account the mass of $s$ quarks.

It was found that the bremsstrahlung ($qq\rightarrow qq\gamma$, $qg\rightarrow qg\gamma$) and annihilation ($qq\bar{q}\rightarrow q\gamma$, $gq\bar{q}\rightarrow g\gamma$) processes corresponding to the 2-loop HTL $\left[O(\alpha\alpha^{2}_{s})\right]$ contribute in the same order of 1-loop HTL $\left[O(\alpha\alpha_{s})\right]$ \cite{th photon 5}. The parametrization of the above 2-loop HTL contributions is used in the present work. However, in Ref.\cite{th photon 5} a numerical error led to an overestimation of the 2-loop HTL rate \cite{th photon 1,th photon 6}, we considered this modification in this paper.

In the QGP phase and mixed phase (MP), the yield of thermal photons produced from the thermal parton interactions can be written as
\begin{eqnarray}
\frac{dN_{\mathrm{QGP}}}{d^{2}p_{T}dy}\!\!\!&&=\!\!\int_{\tau_{0}}^{\tau_{c}}\!\! \!\! \tau d\tau \!\!\int_{0}^{R_{Q}}\!\! \!\! d^{2}r\!\!\int_{-\eta_{\mathrm{lim}}}^{\eta_{\mathrm{lim}}}\!\! \!\! d \eta  E_{\gamma}\!\!\frac{dR_{\mathrm{QGP}}}{d^{3}p} \nonumber\\[1mm]
&&\!\!\!\! +\int_{\tau_{c}}^{\tau_{h}}\!\! \!\! \tau d\tau \!\!\int_{0}^{R_{MP}}\!\! \!\! \!\! d^{2}r\!\!\int_{-\eta_{\mathrm{lim}}}^{\eta_{\mathrm{lim}}}\!\! \!\! d \eta  E_{\gamma}\!\!\frac{dR_{\mathrm{QGP}}}{d^{3}p} f_{\mathrm{QGP}},
\end{eqnarray}
where $y$ and $\eta$ is the rapidity of the photon and fluid element, respectively. The limit of the rapidity is $|\eta_{\mathrm{lim}}|=\mathrm{arcosh}\left(\sqrt{s_{NN}}/2 \mathrm{GeV}\right)$ \cite{th photon 1}.  $R_{Q}=R_{\perp 0}+v_{r}(\tau-\tau_{0})$ and $R_{MP}=R_{\perp 0}+v_{r}(\tau-\tau_{h})$ is the transverse radius of the QGP and MP system, respectively \cite{th photon 8}. The energy of the thermal photon is $E_{\gamma}=p_{T}\cosh(y-\eta)$. $\tau_{c}$ is the critical time when the QGP phase transfers into the mixed phase, and $\tau_{h}$ is the time when the mixed phase transfers into the hadronic phase. The fraction of the QGP matter is $f_{\mathrm{QGP}}=\varepsilon-\varepsilon_{H}/(\varepsilon_{Q}-\varepsilon_{H})$ \cite{1+3D3}.

\subsection{Thermal photons from HG}

In Ref. \cite{th photon 2} Kapusta \emph{et al}. have calculated the differential cross sections for meson annihilation and Compton processes which produce thermal photons: $\pi\pi\rightarrow\rho\gamma$, $\pi\rho\rightarrow\pi\gamma$, $\pi\pi\rightarrow\gamma\gamma$, $\pi\pi\rightarrow\eta\gamma$ and $\pi\eta\rightarrow\pi\gamma$. A rather extended analysis of $\pi\rho a_{1}$-meson gas (including strangeness reactions: $\pi K^{*}\rightarrow K\gamma$, $\pi K\rightarrow K^{*}\gamma$, $\rho K\rightarrow K\gamma$ and $ K K^{*}\rightarrow \pi\gamma$) has studied by Turbide \emph{et al.} \cite{HG photon 1}. The production rate of thermal photons $\pi\rho\rightarrow\pi\gamma$ with $\pi$, $\phi$, $\omega$ and $a_{1}$ mesons as exchange particles for non-strange initial state has been considered \cite{HG photon 3}. The contribution of hadronic bremsstrahlung for the most abundant $\pi\pi\rightarrow\pi\pi\gamma$ channel has also been discussed in Ref. \cite{HG photon 2}. We use the parameterizations of these photon production rates $E_{\gamma}dR_{\mathrm{HG}}/d^{3}p$ in the resent work:
\begin{eqnarray}
\frac{dN_{\mathrm{HG}}}{d^{2}p_{T}dy}\!\!\!&&=\!\!\int_{\tau_{c}}^{\tau_{h}}\!\! \!\! \tau d\tau \!\!\int_{0}^{R_{MP}}\!\! \!\! d^{2}r\!\!\int_{-\eta_{\mathrm{lim}}}^{\eta_{\mathrm{lim}}}\!\! \!\! d \eta  E_{\gamma}\!\!\frac{dR_{\mathrm{HG}}}{d^{3}p}f_{\mathrm{HG}} \nonumber\\[1mm]
&&\!\!\!\! +\int_{\tau_{h}}^{\tau_{f}}\!\! \!\! \tau d\tau \!\!\int_{0}^{R_{H}}\!\! \!\! \!\! d^{2}r\!\!\int_{-\eta_{\mathrm{lim}}}^{\eta_{\mathrm{lim}}}\!\! \!\! d \eta  E_{\gamma}\!\!\frac{dR_{\mathrm{HG}}}{d^{3}p} ,
\end{eqnarray}
where $\tau_{f}$ is the freeze-out time when the temperature reaches the freeze-out temperature of $T_{f}=$120 MeV \cite{th photon 1}. $R_{H}$ is the transverse radius
of HG in the hydrodynamical evolution. The fraction of the HG matter is $f_{\mathrm{HG}}=\varepsilon_{Q}-\varepsilon/(\varepsilon_{Q}-\varepsilon_{H})$ \cite{1+3D3}.

\section{Jet-medium interaction }

\subsection{Jet-photon conversion }
In the relativistic heavy-ion collisions a jet passing through the QGP can interact with a thermal parton. Fries \emph{et al.} have calculated the production of high energy photons from Compton scattering and annihilation of a quark jet passing through the hot medium \cite{jet-photon 1,jet-photon 2}. The production rate of the jet-photon conversion can be written as
\begin{eqnarray}
E_{\gamma}\frac{dR_{\mathrm{\mathrm{jet}-\gamma}}}{d^{3}p}&=&\sum_{q}\left(\frac{e_{q}}{e}\right)^{2}\frac{\alpha\alpha_{s}T^{2}}{8\pi^{2}} \left[f^{\mathrm{jet}}_{q}(\emph{\textbf{p}}_{\gamma}) +f^{\mathrm{jet}}_{\bar{q}}(\emph{\textbf{p}}_{\gamma}) \right]
\nonumber\\[1mm]
&&\times
\left[2\mathrm{ln}\left(\frac{3E_{\gamma}}{\pi\alpha_{s}T}\right)\!+C_{\mathrm{Com}}\!+C_{\mathrm{ann}}\right],
\end{eqnarray}
where $f^{\mathrm{jet}}_{q}$ and $f^{\mathrm{jet}}_{\bar{q}}$ are the phase-space distribution for the quark jets propagating through the hot medium. The sum runs over the $u(\bar{u})$, $d(\bar{d})$, and $s(\bar{s})$ quarks.

In the present work, we consider the contribution of the gluon jet in the $g_{\mathrm{jet}}q_{\mathrm{th}}\rightarrow\gamma q$ interaction:
\begin{eqnarray}
E_{\gamma}\!\frac{dR_{\mathrm{\mathrm{jet}-\gamma}}}{d^{3}p}\!\!&=&\!\!\sum_{q}\left(\frac{e_{q}}{e}\right)^{2}\frac{\alpha\alpha_{s}T^{2}}{6\pi^{2}} f^{\mathrm{jet}}_{g}\!(\emph{\textbf{p}}_{\gamma})\!\!\nonumber\\[1mm]
&&\times
\left[\mathrm{ln}\!\!\left(\!\frac{3E_{\gamma}}{\pi\alpha_{s}T}\right)\!\!+\!C'_{\mathrm{Com}}\!\right],
\end{eqnarray}
where the parameter is $C'_{\mathrm{Com}}=$ 0.046. In this approximation the Fermi-Dirac distribution of thermal quarks is replaced by the Boltzmann distribution due to the large energy of emitted photons \cite{th photon 2}.

The phase-space distribution of jets is given by
\begin{eqnarray}
f^{\mathrm{jet}}_{a}(\emph{\textbf{p}})\label{jet distribution}&=&\frac{(2\pi)^{3}}{g_{a}\pi
R_{\bot}^{2}\tau
p^{jet}_{T}}\frac{dN _{\mathrm{jet}}}{d^{2}p^{\mathrm{jet}}_{T}dy_{\mathrm{jet}}}\delta(\eta-y_{\mathrm{jet}}) \nonumber\\[1mm]
&&\times \Theta(\tau-\tau_{0})
\Theta(\tau_{max}-\tau)\Theta(R_{\bot}-r),
\end{eqnarray}
where $g_{a}$=$2\times 3$ and $2\times 8$ is the spin and color degeneracy of quarks and gluons, respectively. $R_{\bot}$
is the transverse radius of the system. We take $\tau_{max}$ as the smaller of the lifetime of the QGP
and the time taken by the jet produced at position $r$ to reach the
surface of the QGP \cite{jet-photon 1}.

We use the WHS and BDMPS frameworks to calculate the energy loss of the momentum distribution of jets passing through the expanding QGP.
The yield $dN_{\mathrm{jet}}/d^{2}p^{jet}_{T}dy_{jet}$ for producing jets with energy loss in the hot medium can be written as \cite{fu}
\begin{eqnarray}
\frac{dN_{\mathrm{jet}}}{d^{2}p^{\mathrm{jet}}_{T}dy_{\mathrm{jet}}}=C_{n}\!\!\sum_{n=0}^{N}\!\!P_{a}(n)\!\!\left(\!\!1\!\!-\!\!\frac{\triangle E_{a}(n)}{E^{\mathrm{jet}}_{T}}\!\!\right)\!\!\frac{dN^{0}_{\mathrm{jet}}}{d^{2}p^{0 \mathrm{jet}}_{T}\!(n)dy_{\mathrm{jet}}},
\label{jet}
\end{eqnarray}
where $E^{\mathrm{jet}}_{T}(\approx |p^{\mathrm{jet}}_{T}|)$ is the transverse energy of the jet. $p^{0 \mathrm{jet}}_{T}$ is the transverse momentum of the final state parton without the energy loss, we have $p^{0 \mathrm{jet}}_{T}(n)=p^{ \mathrm{jet}}_{T}+\triangle E_{a}(n)$. We discuss the jets produced at midrapidity.

\begin{table}
\centering\caption{Parameters of the jet production yield at $y_{\mathrm{jet}}=0$ for Pb+Pb collisions at $\sqrt{s_{NN}}$=2.76 TeV. The $p^{0\mathrm{jet}}_{T}$ range of the validity is 2 GeV$\leq p^{0\mathrm{jet}}_{T}\leq$20 GeV. (The quarks and gluon have same value of the parameter $h$ for a certain centrality class.)}
\label{temperature}
\begin{tabular}{rcccccccc}
\hline
\hline & & $a$ & $b$ & $c$ & $f$ &  \\
\hline
&          $u$ & $ 0.34444 $&$ 0.20206 $&$ 0.20485 $&$ 1.11097 $&\\
&          $d$ & $ 0.33803 $&$ 0.20255 $&$ 0.20699 $&$ 1.10943 $&\\
& $s(\bar{s})$ & $ 0.45058 $&$ 0.13268 $&$ 0.31029 $&$ 1.69669 $&\\
&    $\bar{u}$ & $ 0.38507 $&$ 0.19017 $&$ 0.20386 $&$ 1.20039 $&\\
&    $\bar{d}$ & $ 0.38392 $&$ 0.18812 $&$ 0.20631 $&$ 1.21430 $&\\
&          $g$ & $ 0.08968 $&$ 0.02582 $&$ 0.34654 $&$ 1.91486 $&\\

\hline
\hline
         &Centrality & $0-20\% $&$ 20-40\% $&$ 40-80\%$ & \\
\hline
& $h$[1/GeV$^{2}$] &$ $ $0.29134 $&$ 0.15635 $&$ 0.01564$ &\\

\hline
\end{tabular}
\end{table}
\begin{figure}[t]
\begin{center}
\includegraphics[width=11 cm]{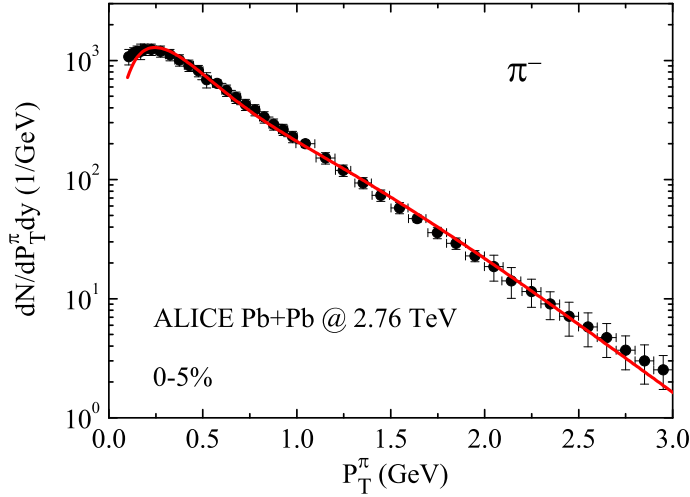}
\caption{\label{pion}  (Color online) Pion yield from the Pb+Pb collisions at $\sqrt{s_{NN}}$=2.76 TeV for 0-5\% central collisions. Solid line denotes the numerical results of the blast-wave parameterization. }
\end{center}
\end{figure}

\begin{figure}[t]
\begin{center}
\includegraphics[width=11 cm]{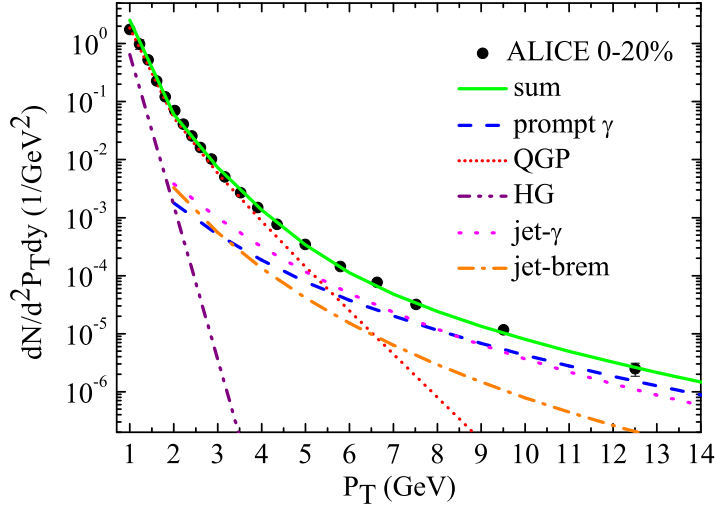}
\caption{\label{photon1}  (Color online) Photon yield from the Pb+Pb collisions at $\sqrt{s_{NN}}$=2.76 TeV for 0-20\% centrality class. The initial temperature and time is $T_{0}$=762 MeV and $\tau_{0}$=0.15 fm/$c$, respectively. Dash line: prompt photons produced from the cold components interactions (direct and fragmentation processes). Short dot line: thermal photons from the QGP. Dash dot dot line: thermal photons from the HG. Dot line: photons produced from the jet-photon conversion. Dash dot line: photons from the jet bremsstrahlung in the medium. Solid line: the sum of the above contributions. The data of direct photons is from the ALICE experiments \cite{ALICE 1}. }
\end{center}
\end{figure}

\begin{figure}[t]
\begin{center}
\includegraphics[width=11 cm]{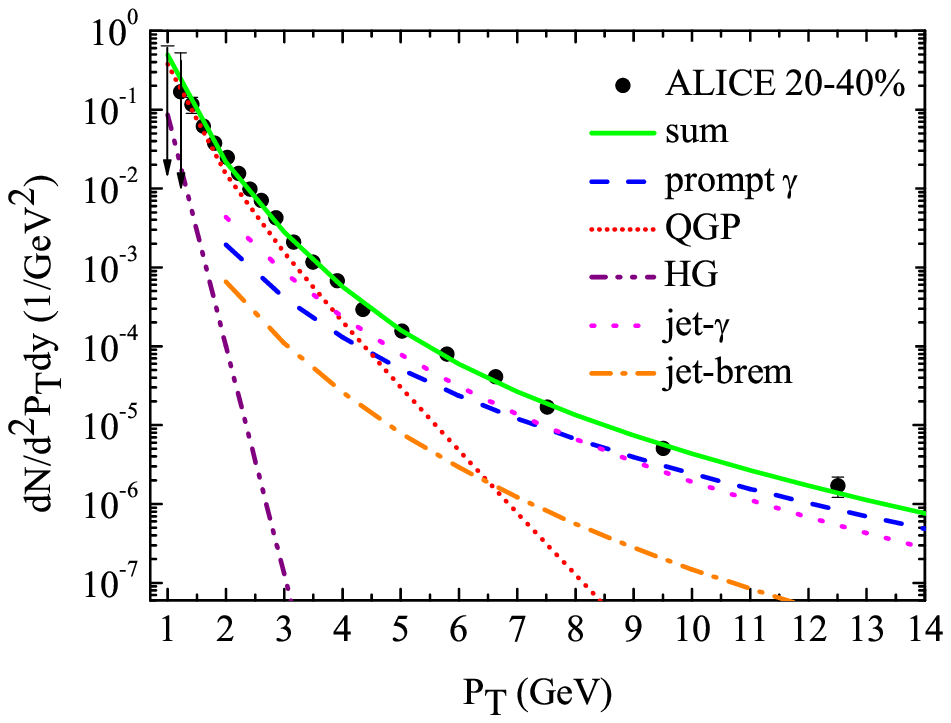}
\caption{\label{photon2}  (Color online) Same as Fig.\ref{photon1} but for the 20-40\% centrality class, $T_{0}$=680 MeV and $\tau_{0}$=0.15 fm/$c$. }
\end{center}
\end{figure}

The jet production yield $dN^{0}_{\mathrm{jet}}/d^{2}p^{0\mathrm{jet}}_{T}dy_{\mathrm{jet}}$ without the energy loss in the relativistic heavy-ion collisions ($A+B\rightarrow \mathrm{jet}+X$) can be factored in the pQCD theory as
\begin{eqnarray}\label{eq8}
\frac{dN^{0}_{\mathrm{jet}}}{d^{2}p^{0\mathrm{jet}}_{T}dy_{\mathrm{jet}}}&\!\!\!\!=\!\!\!&\frac{\langle N_{\mathrm{coll}} \rangle}{\sigma^{NN}_{\mathrm{inel}}}\sum_{a,b}
\!\int_{x_{a}^{min}}^{1}\!\!dx_{a}f^{A}_{a}(x_{a},Q^{2})f^{B}_{b}(x_{b},Q^{2}) \nonumber\\[1mm]
&& \times\frac{x_{a}x_{b}}{\pi(x_{a}-x_{1})}K_{\mathrm{jet}}\frac{d\hat{\sigma}_{ab\rightarrow
cd}}{d\hat{t}}(\hat{s},\hat{u},\hat{t}),
\end{eqnarray}
where the momentum fractions with the rapidity
are given by $x_{a}^{min}=x_{1}/(1-x_{2})$ and $x_{b}=x_{a}x_{2}/(x_{a}-x_{1})$,
here the variables are $x_{1}=x_{T}e^{y_{\mathrm{jet}}}/2$, $x_{2}=x_{T}e^{-y_{\mathrm{jet}}}/2$,
$x_{T}=2p^{0\mathrm{jet}}_{T}/\sqrt{s_{NN}}$. $K_{\mathrm{jet}}=$1.6 is the NLO pQCD
correction factor \cite{jet-photon 3}. We also give the parametrization of the jet yield for the simplicity of the numerical calculations:
\begin{eqnarray}
\frac{dN^{0}_{\mathrm{jet}}}{d^{2}p^{0\mathrm{jet}}_{T}dy_{\mathrm{jet}}}=K_{\mathrm{jet}}h\left[a+b\left(\frac{p^{0\mathrm{jet}}_{T}}{1 \mathrm{GeV}}\right)^{f}\right]^{-1/c}_{y_{\mathrm{jet}}=0},
\end{eqnarray}
the numerical values of the parameters $a$, $b$, $c$, $f$, and $h$ are listed in Table II.

\subsection{Jet bremsstrahlung}
In Ref. \cite{jet-brem 1} Zakharov has discussed the induced bremsstrahlung from a charged jet due to multiple scattering in the QGP. If we assume $y=y_{\mathrm{jet}}$=0, the yield of photons produced from the jet bremsstrahlung in the QGP can be written as
\begin{eqnarray}
\frac{dN_{\mathrm{jet-brem}}}{d^{2}p_{T}dy}=\int^{1}_{0}\frac{dx}{x^{2}}\frac{dP_{ind}(x,p_{T}^{\mathrm{jet}})}{dx}\frac{dN_{\mathrm{jet}}}{d^{2}p^{\mathrm{jet}}_{T}dy_{\mathrm{jet}}},
\end{eqnarray}
where the momentum fraction is $x=p_{T}/p_{T}^{\mathrm{jet}}$, and the jet production yield is from Eq.(\ref{jet}). The radiation rate of the bremsstrahlung without the energy loss is
\begin{eqnarray}
\frac{dP^{0}_{ind}(x,p_{T}^{0\mathrm{jet}})}{dx}&=&\left(\sum_{q}\frac{e_{q}}{e}\right)^{2}\pi\alpha\alpha_{s}C_{T}C_{F}n_{m}L^{2},
\nonumber\\[1mm]
&&\times \frac{1\!-\!x\!+\!x^{2}/2}{8p_{T}^{0\mathrm{jet}}(1-x)}
\end{eqnarray}
where $C_{T}$=3 and $C_{F}$=4/3 is the color Casimir for the medium constituents and quark jet, respectively. $n_{m}=\frac{T^{3}}{2\pi^{2}}\left[15\times\frac{3}{2}\xi(3)+16\times\Gamma(3)\xi(3)\right]$ is the number density of the medium \cite{th photon 4}. The effective radiation rate considering the jet energy loss can be written in the following
\begin{eqnarray}
\frac{dP_{ind}(x,p_{T}^{\mathrm{jet}})}{dx}=C_{n}\sum^{N}_{n=0}P_{a}(n)\frac{x^{n}}{x}\frac{dP^{0}_{ind}(x,p_{T}^{0\mathrm{jet}}(n))}{dx},
\end{eqnarray}
here $x^{n}=x/\left[1-(\triangle E_{a}/E_{T}^{0\mathrm{jet}})\right]$. The jet travels only a short distance and scatter a few times through the plasma in the jet-medium interactions \cite{jet-photon 2}. In the high energy limit, the spectrum of the photon bremsstrahlung is dominated by the $n$=1 scattering due to the finite size effects. The distance is limited as $L\sim \lambda_{a}< L_{f}^{\gamma}$, where $L_{f}^{\gamma}$ is the photon formation length \cite{jet-brem 1}.

\begin{figure}[t]
\begin{center}
\includegraphics[width=11 cm]{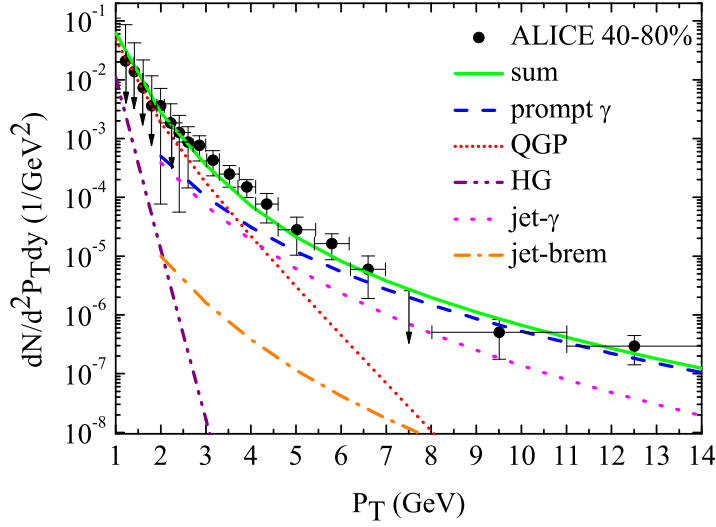}
\caption{\label{photon3}  (Color online) Same as Fig.\ref{photon1} but for the 40-80\% centrality class, $T_{0}$=538 MeV and $\tau_{0}$=0.15 fm/$c$. }
\end{center}
\end{figure}

\section{Numerical results and discussion}

The hydrodynamic initial conditions are fixed by the hadron multiplicity. We also compare the transverse momentum spectrum of charged pion to data from Pb+Pb (0-5\%) central collisions at $\sqrt{s_{NN}}$=2.76 TeV \cite{pion data} in Fig.\ref{pion}. We calculate the spectrum by using the blast-wave parameterization \cite{blast wave} which is considered the (1+1)-D space-time evolution. One can see that the numerical result can reproduce the data in the low $p_{T}^{\pi}$ region where the thermal production and resonance decays are expected to be the main mechanism of pion production.

From Fig.\ref{photon1}-Fig.\ref{photon3} we plot the contributions of photons produced from the pQCD interactions, thermal parton and hadron scattering, and high energy jet-medium interactions in the Pb+Pb collisions at $\sqrt{s_{NN}}$=2.76 TeV. The numerical results agree with the ALICE experimental data of the direct photons \cite{ALICE 1}. The ALICE Collaboration has studied the direct photon production at mid-rapidity in Pb+Pb collisions at $\sqrt{s_{NN}}$=2.76 TeV. The direct photons are defined as photons not originating from hadron decays: $\gamma_{\mathrm{direct}}=\gamma_{\mathrm{incl}}-\gamma_{\mathrm{decay}}$, where $\gamma_{\mathrm{incl}}$ is the measured inclusive photon spectrum, and $\gamma_{\mathrm{decay}}$ is the decay photon spectrum.

The numerical results indicate that the prompt photons produced from the pQCD hard scattering are dominant in the large transverse momentum region of $p_{T}>$ 9.5 GeV (for Fig.\ref{photon1} 0-20\% centrality), 8.5 GeV (for Fig.\ref{photon2} 20-40\% centrality), and 3.7 GeV (for Fig.\ref{photon3} 40-80\% centrality), respectively. The nuclear modification and energy loss of jet fragmentation are considered in the calculations. Photons from the QGP, HG and jet-medium interactions are dominant below these regions.

The QGP is an important photon production source in the relatively small $p_{T}$ region. The numerical results of Fig.\ref{photon1}, Fig.\ref{photon2} and Fig.\ref{photon3} show that the thermal photons from the QGP are dominant in the transverse momentum region of $p_{T}<$ 5 GeV (for $T_{0}$=762 MeV), 4 GeV (for $T_{0}$=680 MeV), and 3.5 GeV (for $T_{0}$=538 MeV), respectively. In this paper, we study the (1+1)-D relativistic hydrodynamic evolution. We take the initial time to be $\tau_{0}$=0.15 fm/$c$ with initial temperatures $T_{0}$=762, 680, and 538 MeV for the 0-20\%, 20-40\%, and 40-80\% centrality classes, respectively. The hydrodynamics equation (\ref{hydro}) is solved numerically using the first-order Lax finite difference scheme \cite{1+3D5}. In the numerical calculations, the numerical viscosity effects have been shown to be negligible for the lattice spacings and time steps. We also compare with the results of earlier works \cite{1+3D1,1+3D3,1+3D4} to ensure that technical aspects are under control. We evaluate the critical time of the longitudinal and transverse expanding QGP as $\tau_{c}$=10.8 fm/$c$ (for $T_{0}$=762 MeV), $\tau_{c}$=8.6 fm/$c$ (for $T_{0}$=680 MeV), and $\tau_{c}$=5.1 fm/$c$ (for $T_{0}$=538 MeV). For comparison, the critical time of the Bjorken expansion is given as $\tau_{c}^{\mathrm{Bjor}}=\tau_{0}(T_{0}/T_{c})^{3}$. We have $\tau_{c}^{\mathrm{Bjor}}$=16.2, 11.5, and 5.7 fm/$c$ for $T_{0}$=762, 680, and 538 MeV, respectively.. We find that the life time ($\triangle\tau=\tau_{c}-\tau_{0}$) of the (1+1)-D expanding QGP is smaller than the life time of the 1-D Bjorken expansion. The transverse flow leads to a rapid cooling of the fire ball.

We find that the thermal photons from the QGP shine more bright than the HG. The spectra of the thermal photons from the HG fall off with the transverse momentum of photons faster than the spectra of thermal photons produced by QGP due to $T_{\mathrm{HG}}<T_{\mathrm{QGP}}$ and $\varepsilon_{\mathrm{HG}}\ll\varepsilon_{\mathrm{QGP}}$. The transverse flow effect also leads to a more rapid cooling of the HG and a reduced life time of the interacting hadronic gas. The contribution of the HG is found to be small in Fig.\ref{photon1}-Fig.\ref{photon3}, just about 20\%$\sim$30\% at $p_{T}\sim$1 GeV for Pb+Pb $\sqrt{s_{NN}}$=2.76 TeV collisions.

The jet-photon conversion in the medium is also an important photon production source. The jet-photon conversion includes the interaction of thermal and cold components. Since the rate of the jet-photon conversion is $dR_{\mathrm{jet-\gamma}}/d^{2}p_{T}dy\propto f_{\mathrm{jet}}$, the spectra of the jet-photon conversion do not drop quickly with the transverse momentum. In Fig.\ref{photon1} and Fig.\ref{photon2} the jet-photon conversion is dominant in the region of 5.2 GeV$<p_{T}<$7.5 GeV (for 0-20\% centrality and $T_{0}$=762 MeV), 4 GeV$<p_{T}<$7 GeV (for 20-40\% centrality and $T_{0}$=680 MeV). However, in Fig.\ref{photon3} the spectrum of the jet-photon conversion is covered by the spectrum of prompt photons in the 40-80\% centrality class due to the centrality dependence of the jet distribution $f_{\mathrm{jet}}$ and the system temperature \cite{jet-photon 2}. In Ref. \cite{jet-photon 1,jet-photon 5} the jet distribution was approximated as $\bar{f}_{\mathrm{jet}}=\sum_{q}e_{q}^{2}f^{q}_{\mathrm{jet}}/\sum_{q}e_{q}^{2}$, the contribution of gluon jet was also to be neglected. In this paper we consider all light quark jets and gluon jet. We find that the contribution of gluon jet in the jet-photon conversion is suppressed by the factor $1/g_{a}$ ($g_{a}=6$ for the light quarks and 16 for the gluon) in Eq.(\ref{jet distribution}).

In Fig.\ref{photon1} the photon spectrum of the jet bremsstrahlung in the medium is smaller than the spectrum of the jet-photon conversion for 0-20\% centrality class. Between 4 and 10 GeV, the contribution of the jet bremsstrahlung is approximately 10\% to that of the total photon spectrum. The photon production via jet bremsstrahlung in the medium turns out to be weak for 20-40\%(Fig.\ref{photon2}) and 40-80\%(Fig.\ref{photon3}) centrality classes.

The jet energy loss is also included in the jet-medium interactions. The energy loss $dE/dx$ is proportional to the distance that the jet has raveled in the QGP. We assume the jet is massless and travels with the speed of light in the transverse direction. If an energetic jet paces through a long distance in the QGP, and fragments outside the system, the average distance is $\langle L \rangle\approx 0.9 R_{\bot}$, the energy loss of the jet is large \cite{WHS1,WHS2,WHS3,WHS4}. However, in the case of the jet-medium interaction, jets travel a relatively short distance through the plasma before they convert into photons, and do not lose a significant amount of energy \cite{jet-photon 2,fu}.

\section{Summary}
We investigate the production of photons from the hard scattering of the cold components, the thermal interactions of the QGP and HG, and the jet-medium interactions in the Pb+Pb collisions at $\sqrt{s_{NN}}$=2.76 TeV. The jet energy loss of the photon fragmentation function, the jet distribution, and the photon radiation rate of the bremsstrahlung is included in the calculations by using the WHS and BDMPS models. We use the (1+1)-D ideal relativistic hydrodynamics to study the space-time evolution of the hot medium. The transverse flow leads to the decreasing of the life time of the QGP and HG. We find that the prompt photons from the initial parton scattering, thermal photons from the QGP, and photons from the jet-photon conversion are the dominant photon production sources in Pb+Pb collisions at LHC energy. The sum of spectra is compared with the ALICE data of the direct photons from the Pb+Pb collisions at $\sqrt{s_{NN}}$=2.76 TeV. The agreement with data is good for the 0-20\%, 20-40\% and 40-80\% centrality classes.

\section{Acknowledgements}
This work is supported by the Applied Fundamental Research Program (AFRP) of Yunnan Province of
China under Grant No. 2017FD250.





\bibliographystyle{model1-num-names}
\bibliography{<your-bib-database>}







\end{document}